\documentclass[aps,prd,twocolumn,floats,nofootinbib,longbibliography]{revtex4-1}
\usepackage[dvips]{graphicx}

\usepackage{amsmath}
\usepackage{braket}
\usepackage{amsfonts}
\usepackage[colorlinks=true]{hyperref}
\usepackage{hyperref}
\usepackage{xcolor}
\usepackage{amssymb}

\usepackage{booktabs}

\usepackage{aas_macros}

\usepackage{caption}
\usepackage{graphicx}

\usepackage{bm}

\usepackage{changes}
\definechangesauthor[name={Olivier}, color=magenta]{OM}
\definechangesauthor[name={Denis}, color=teal]{DA}

\def\aap{\textit{Astronomy and Astrophysics}}
\def\dd {\mbox{d}}

\def\be{\begin{equation}}
\def\ee{\end{equation}}
\def\bea{\begin{eqnarray}}
\def\eea{\end{eqnarray}}

\def\Lm{\mathcal{L}_m}
\def\L{\mathcal{L}}

\def\varphip{\varphi / \sqrt{3}}

\def\t{\tilde}

\begin{document}
%%-----------------------------
%%      the top matter
%%-----------------------------
\title{Analytical external spherical solutions in entangled relativity}

%\author[0000-0002-3151-7593]{Olivier Minazzoli}
\author{Denis Arruga}
%\affiliation{Laboratoire Astroparticule et Cosmologie (APC), Universit\'e de Paris, France}
%\affiliation{???}
\author{Olivier Minazzoli}
%\affiliation{Centre Scientifique de Monaco, 8 Quai Antoine 1er, 98000, Monaco}
\affiliation{Artemis, Universit\'e C\^ote d'Azur, CNRS, Observatoire C\^ote d'Azur, BP4229, 06304, Nice Cedex 4, France}

\begin{abstract}
In this manuscript, we present analytical external spherical solutions of entangled relativity, which we compare to numerical solutions obtained in a Tolman-Oppenheimer-Volkoff framework. Analytical and numerical solutions match perfectly well outside spherical compact objects, therefore validating both types of solutions at the same time. The analytical external (hairy) solutions---which depend on two parameters only---may be used in order to easily compute observables---such as X-ray pusle profiles---without having to rely on an unknown equation of state for matter inside the compact object.

\end{abstract}
\maketitle

%--------------------------------------------------
%\section*{}
\section{Introduction}

Entangled relativity is a new general theory of relativity that changes the way spacetime and matter interract with each other \cite{ludwig:2015pl,minazzoli:2018pr,arruga:2021pr}.\footnote{The name \textit{entangled relativity} appears for the first time in \cite{arruga:2021pr}.} Instead of assuming that the spacetime and matter parts of the action have to be glued together additively, it is assumed in entangled relativity that they are glued together multiplicatively instead. This has the immediate consequence that gravity and inertia cannot be defined without defining matter at the same time, and vice-versa---therefore satisfying Einstein's main version of Mach's Principle  \cite{einstein:1918an}\footnote{A translation in English of the original paper in German is available online at \href{https://einsteinpapers.press.princeton.edu/vol7-trans/49}{https://einsteinpapers.press.princeton.edu/vol7-trans/49}.} \cite{pais:1982bk}.

While the pure mutliplicative coupling in the action could (na\"ively) question the viability of the theory, it turns out that the action can be written in a form of a scalar-tensor theory that possesses an \textit{intrinsic decoupling} of the scalar-field degree of freedom \cite{ludwig:2015pl,minazzoli:2018pr}. This means that the theory possesses the same degrees of freedom as a scalar-tensor theory---therefore ensuring its theoretical viability---and that the scalar-field is not, or weakly, sourced in most situations, such that the phenomenology of entangled relativity is very close to the one of general relativity in many cases \cite{minazzoli:2013pr,minazzoli:2014pr,minazzoli:2014pl,minazzoli:2020ds,minazzoli:2020bh,arruga:2021pr}. 

Recently, we studied numerical solutions of compact objects in entangled relativity \cite{arruga:2021pr}. Here, we present analytical external solutions for spherical objects, which we then compare to our numerical solutions. We find that the analytical and numerical solutions match with each other, providing evidence of the validity of each of them. Since the analytical solutions depend on two paremeters only, they may be usefull in order to easily compute observables related to neutron stars---such as X-ray pulse profiles \cite{riley2019:ap,miller:2019ap,bogdanov:2019ap}; whereas numerical solutions could then  be used, in a second time, in order to check what types of equation of state can produce the fitted values of the parameters. 

Along the way, we recovered an old analytical solution for spherical objects with scalar hairs that does not seem to be widely known by the community.

%The analytical solutions possess scalar hairs, which are parametrized by a parameter

%--------------------------------------------------
\section{Field equations}
\label{sec:field}

The action of entangled relativity is defined by \cite{minazzoli:2018pr,ludwig:2015pl}
\be
\label{eq:actER}
S=-\frac{\xi}{2} \int \mathrm{d}^{4} x \sqrt{-g} \frac{\mathcal{L}_{m}^{2}}{R},
%S=-\frac{\xi}{2} \int \mathrm{d}_g^{4} x  \frac{\mathcal{L}_{m}^{2}}{R},
\ee
where
% $\mathrm{d}_g^{4} x \equiv \mathrm{d}^{4} x \sqrt{-g}$ is the space-time volume element,
the coupling constant $\xi$ has the dimension of $\kappa \equiv 8 \pi G / c^{4}$---where $G$ is the Newtonian constant and $c$ the speed of light---but not its value. In fact, $\xi$ does not appear in the field equations that derive from the extremization of the action (\ref{eq:actER}), and therefore is purely related to the quantum field sector of the theory. It is important to note that apart from $\xi$, the theory does not have any coupling constant related to the link between matter and geometry. Hence, at the classical level, entangled relativity has one parameter less than general relativity in order to describe the link between matter and geometry, in the sense that no parameter replaces the parameter $\kappa$ of general relativity at the classical level in entangled relativity \cite{minazzoli:2018pr}: the effective coupling that appears at the level of the field equation is dynamical. For $\Lm \neq 0$, the metric field equation reads\footnote{The metric field equation for all $\Lm$ is given in the Appendix \ref{app:fe}.}
\be
R_{\mu \nu}-\frac{1}{2} g_{\mu \nu} R= - \frac{R}{\Lm} T_{\mu \nu} + \frac{R^2}{\Lm^2} \left(\nabla_\mu \nabla_\nu - g_{\mu \nu} \Box \right)\frac{\Lm^2}{R^2}, \label{eq:fRmetricfield}
\ee
with
\be
T_{\mu \nu} \equiv-\frac{2}{\sqrt{-g}} \frac{\delta\left(\sqrt{-g} \mathcal{L}_{m}\right)}{\delta g^{\mu \nu}}.
\ee
Also note that the stress-energy tensor is no longer conserved in general, as one has
\be
\nabla_{\sigma}\left(\frac{\mathcal{L}_{m}}{R} T^{\alpha \sigma}\right)=\mathcal{L}_{m} \nabla^{\alpha}\left(\frac{\mathcal{L}_{m}}{R}\right). \label{eq:noconsfR}
\ee
Otherwise, note that the trace of Eq. (\ref{eq:fRmetricfield}) reads
\be
3 \frac{R^{2}}{\mathcal{L}_{m}^{2}} \square \frac{\mathcal{L}_{m}^{2}}{R^{2}}=-\frac{R}{\Lm}\left(T-\Lm\right). \label{eq:fRmetricfieldt}
\ee
The \textit{intrinsic decoupling} of the scalar degree of freedom is manifest for $\Lm = T$ \cite{minazzoli:2013pr,minazzoli:2014pr,minazzoli:2014pl,minazzoli:2016pr,minazzoli:2018pr,minazzoli:2020ds}. Indeed, for $\Lm = T$ on-shell---such as for a dust field, or null-radiation---$\L_m/R$ is solution of the trace of the metric field equation (\ref{eq:fRmetricfieldt}), such that one recovers the metric field equation of general relativity in that case \cite{minazzoli:2018pr}. 

It is important to notice that the coupling constant between matter and geometry in the metric field equation of general relativity is replaced by a scalar field degree of freedom in entangled relativity $8 \pi G_{\textrm{eff}}/c^4 := - R/\Lm$ \cite{minazzoli:2018pr}. In particular, the effective coupling in the metric field equation of entangled relativity is positive for $\Lm /R < 0$ and negative for $\Lm /R > 0$, potentially providing a repulsive mechanism at high density, where the kinetic energy density should dominate in the on-shell matter Lagrangian \cite{minazzoli:2021bh,minazzoli:2021dm}. \footnote{While the transition between the attractive and repulsive cases seems to be singular in Eq. (\ref{eq:fRmetricfield}), it is not the case when one looks at the actual metric field equation for all $\Lm$ given in Eq. (\ref{eq:fRmetricfield_o}).}

Indeed, as a star collapse into a black hole, the kinetic part $K$ of the matter Lagrangian density shall ineluctably start to dominate the whole Lagrangian density on-shell---that is $K>V$, such that $\Lm := K-V>0$, where $V$ is the potential part of the Lagrangian density. If, in the meantime, $R$ keeps the same sign, then gravity shall become repulsive and the collapsing matter shall rebound due to the new repulsive nature of gravity. This should avoid the formation of spacetime singularities inside black holes in the framework of entangled relativity. 

The avoidance of singularities in entangled relativity is somewhat expected given that the theory prohibits the existence of spacetime without matter at a fundamental level \cite{ludwig:2015pl,minazzoli:2018pr,minazzoli:2021bh}.

%--------------------------------------------------
\section{Almost equivalent action}

For spacetimes that are such that $(R,\Lm) \neq 0$, there is a one to two correspondence at the classical level between the action of entangled relativity and a dilaton theory with the following action \cite{ludwig:2015pl,minazzoli:2018pr}

\be
\label{eq:sfaction}
S=\frac{1}{c} \frac{\xi}{\tilde \kappa} \int d^{4} x \sqrt{-g}\left[\frac{\phi R}{2 \tilde \kappa}+\sqrt{\phi} \mathcal{L}_{m}\right],
%S=\frac{1}{c} \frac{\xi}{\tilde \kappa} \int d_g^{4} x\left[\frac{\phi R}{2 \tilde \kappa}+\sqrt{\phi} \mathcal{L}_{m}\right],
\ee
where $\tilde \kappa$ is an effective coupling constant between matter and geometry, with the dimension of the coupling constant of general relativity $\kappa$. Because entangled relativity leads to a repulsive gravitational phenomenon for $\Lm / R > 0$, the action in Eq. (\ref{eq:sfaction}) corresponds to the original action in Eq. (\ref{eq:actER}) as long as one has $\tilde \kappa > 0$ if $\mathcal{L}_m/R < 0$, and $\tilde \kappa < 0$ if $\mathcal{L}_m/R > 0$. The corresponding field equations read

\bea
&&G_{\alpha \beta}=\tilde \kappa \frac{T_{\alpha \beta}}{\sqrt{\phi}}+\frac{1}{\phi}\left[\nabla_{\alpha} \nabla_{\beta}-g_{\alpha \beta} \square\right] \phi,\label{eq:metr}\\
%&&\frac{3}{\phi} \square \phi=\frac{\tilde \kappa}{\sqrt{\phi}}\left(T-\mathcal{L}_{m}\right), \label{eq:sceq}
&&\sqrt{\phi}=-\tilde \kappa \mathcal{L}_{m} / R. \label{eq:defphi}
\eea
%\bea
%&&G_{\alpha \beta}=-\tilde \kappa \frac{T_{\alpha \beta}}{\sqrt{\phi}}+\frac{1}{\phi}\left[\nabla_{\alpha} \nabla_{\beta}-g_{\alpha \beta} \square\right] \phi,\label{eq:metrm}\\
%%&&\frac{3}{\phi} \square \phi=\frac{\tilde \kappa}{\sqrt{\phi}}\left(T-\mathcal{L}_{m}\right), \label{eq:sceq}
%&&\sqrt{\phi}=\bar{\kappa} \mathcal{L}_{m} / R, \label{eq:sceqm}
%\eea
%if $\mathcal{L}_m/R > 0$.
 The conservation equation reads
\be
\label{eq:noncons}
\nabla_{\sigma}\left(\sqrt{\phi} T^{\alpha \sigma}\right)=\mathcal{L}_{m} \nabla^{\alpha} \sqrt{\phi},
\ee
%with
%\be
%T_{\mu \nu} \equiv-\frac{2}{\sqrt{-g}} \frac{\delta\left(\sqrt{-g} \mathcal{L}_{m}\right)}{\delta g^{\mu \nu}}.
%\ee
The trace of the metric field equation can therefore be rewritten as follows
\be
\frac{3}{\phi} \square \phi=\frac{\tilde \kappa}{\sqrt{\phi}}\left(T-\mathcal{L}_{m}\right). \label{eq:sceq}
\ee
The equivalence between Eqs. (\ref{eq:metr}-\ref{eq:sceq}) and (\ref{eq:fRmetricfield}-\ref{eq:noconsfR}) is pretty straightforward to check.

%--------------------------------------------------
\section{Generic external vacuum solutions in scalar-tensor theories}
\label{sec:generic}

Let us consider the following generic class of actions\footnote{For now on, we use the unit system that is such that $G=c=\mu_0=1$.}
\be \label{eq:actioncl}
%S=\int d^{4} x \sqrt{-g}\left[R-(\nabla \varphi)^{2} + f(\varphi,\Lm) \right],
S=\int d^{4} x \sqrt{-g}\left[R-2(\nabla \varphi)^{2} + f(\varphi,\Lm) \right],
\ee
that is such that one has in the vacuum limit ($\Lm \rightarrow 0$)
\be \label{eq:const}
\lim_{\Lm \rightarrow 0} f(\varphi,\Lm) \rightarrow 0.
\ee
For instance, Brans-Dick theory in the Einstein frame implies that $f(\varphi,\Lm) = \Lm(A^2(\varphi) g_{\mu \nu})$ \cite{will:2014lr}, whereas low-energy string, supergravity, Kaluza-Klein and entangled relativity theories in the Einstein frame generically imply that $f(\varphi,\Lm) = f(\varphi) \Lm(A^2(\varphi) g_{\mu \nu})$ \cite{damour:1994np,gibbons:1988np,minazzoli:2021bh}, while general relativity corresponds to $f(\varphi,\Lm) = \Lm(g_{\mu \nu})$. 
The action (\ref{eq:sfaction}) can be put in the form of the action (\ref{eq:actioncl}) after the conformal transformation of the metric $g_{\alpha \beta} \rightarrow e^{-2\varphip} g_{\alpha \beta}$, with $\phi = e^{-2\varphip}$---see Sec. \ref{sec:EF}.

We found that a class of vacuum spherical solutions for the generic class of actions (\ref{eq:actioncl}) reads
\bea
ds^2 &=& - \left(1-\frac{2m}{\beta r} \right)^\beta \dd t^2 +\left(1-\frac{2m}{\beta r} \right)^{-\beta} \dd r^2 \nonumber \\
&&+ r^2~\left(1-\frac{2m}{\beta r} \right)^{1-\beta} \left[\dd \theta^2 + \sin^2 \theta ~\dd \psi^2 \right], \label{eq:metric}
\eea
with
\be
%e^\varphi = \left(1-\frac{2m}{\beta r} \right)^{\frac{1-\beta}{2\alpha}} \label{eq:SFm}
\varphi = \frac{1-\beta}{2\alpha}~\ln \left(1-\frac{2m}{\beta r} \right) \label{eq:SFm}
\ee
where 
\be
%\beta := \frac{1-\alpha^2}{1+\alpha^2}. \label{eq:beta}
\alpha^2 = \frac{1-\beta }{1+\beta},\label{eq:beta}
\ee
where $\beta \in ~[-1;0[~ \cup~ ]0; 1]$ (or equivalently $\alpha \in \mathbb{R} - \{1\}$).\footnote{We check these solutions in Appendix \ref{app:equations}.} One can either have $\alpha \geq 0$ or $\alpha \leq 0 $ because the action is invariant under the reflection ($\mathbb{Z}$-2) symmetry $\varphi \rightarrow -\varphi$ at the limit $\Lm =0$. One recovers the usual Schwarzschild metric for $\beta = 1$ (or equivalently $\alpha =0$).\footnote{Actually, one can show that the metric (\ref{eq:metric}) is invariant under the transformation $\beta \rightarrow -\beta$, with a shifted radial coordinate $\rho$ as $\rho = r -2m/\beta$---see Appendix \ref{sec:beta}.} Hence, Eqs. (\ref{eq:actioncl}-\ref{eq:const}) are a generalization of the Schwarzschild metric for all the theories that can be written in the form of Eqs. (\ref{eq:actioncl}-\ref{eq:const}). The solutions described by the Eqs. (\ref{eq:metric}-\ref{eq:beta}) are also much simpler than the Janis-Newman-Winicour solutions \cite{janis:1968pr}---although they ought to describe the same spacetimes, albeit with different coordinates. After some investigation of the literature, we found that the solutions in Eqs. (\ref{eq:metric}-\ref{eq:beta}) can already be found in \cite{damour:1992cq}, and that it has been attributed to Just \cite{just:1959zn}.

It is crucial to understand that Eqs. (\ref{eq:metric}-\ref{eq:beta}) are vacuum solutions of Eqs. (\ref{eq:actioncl}-\ref{eq:const}) for all $\beta \in ~]-1;0[~ \cup~ ]0; 1]$.
It means that any theory that can be written in the form of Eqs. (\ref{eq:actioncl}-\ref{eq:const}), has a family of solutions that reads as Eqs. (\ref{eq:metric}-\ref{eq:beta}), with various values for the parameters $m$ and $\beta$. As a consequence, all theories that can be written as Eqs. (\ref{eq:actioncl}-\ref{eq:const}) have spherical solutions that do not only depend on a mass, but on the parameter $\beta$ as well. We shall therefore qualify these solutions as \textit{hairy} ones.

It is therefore quite different from the charged spherical solutions in the Einstein-Maxwell-dilaton theories \cite{garfinkle:1991pr,holzhey:1992nb}, for which a similar parameter $\alpha$ is fixed by the theory that one considers---like for instance $\alpha =1$ for the tree-level low-energy limit of string theory and ($D=4, N=4$) supergravity, or $\alpha = \sqrt{3}$ for 5D Kaluza-Klein theory \cite{gibbons:1988np}, or $\alpha = 1/(2\sqrt{3})$ for entangled relativity \cite{minazzoli:2021bh}. Here, on the other hand, the theory does not constrain the value of $\alpha$ nor $\beta$. Indeed, $\alpha$ and $\beta$ do not appear in the Lagrangian density, unlike in the Einstein-Maxwell-dilaton cases, where $\alpha$ corresponds to the coupling strength between the scalar field and matter in the Lagrangian density \cite{garfinkle:1991pr,holzhey:1992nb}.

It is important to note that Eqs. (\ref{eq:metric}-\ref{eq:beta}) are solutions of general relativity as well, provided that there is a massless canonical scalar field---for which, as far as we know, there is currenlty no evidence for in nature. One might therefore think that the solutions in Eqs. (\ref{eq:metric}-\ref{eq:beta}) violate Birkhoff's theorem \cite{griffiths:2012bk}, but that is not the case. Indeed, the presence of the scalar-field in the action (\ref{eq:actioncl}) implies that one is not dealing with general relativity in vacuum, while Birkhoff's theorem applies to general relativity in vacuum \cite{griffiths:2012bk}.

It is also important to note that $r= 2m / \beta$ is a curvature singularity for all $\beta \in ~]0; 1[$ (or equivalently $\alpha \in ~ ]0; 1[$)---see Appendix \ref{app:sing}---whereas it is an event horizon for the Schwarzschild case---that is, for $\beta =1$ (or equivalently $\alpha = 0$). However, such a singularity is not expected to happen in nature due to the fact that scalar hairs are radiated away during the collapse into a black-hole in scalar-tensor theories of the form of (\ref{eq:actioncl}-\ref{eq:const}), leading to black holes with no hair \cite{hawking:1972cm,scheel:1995pr,sotiriou:2012pl,graham:2014pr,berti:2015cq}. Therefore, while the solutions in Eqs. (\ref{eq:metric}-\ref{eq:beta}) should be exact external solutions of spherical objects, they should not correspond to pure vacuum solutions---unlike the Schwarzschild and Kerr black holes for instance.

Nevertheless, direct observations of the \textit{shadow} of diverse black holes \cite{mizuno:2018na,eht:2019aj}---such as the one done with the Event Horizon Telescope for M87 \cite{eht:2019aj}---and the corresponding signatures of photons \textit{subrings} \cite{johnson:2020sc}, could be used in order to test $\beta \neq 1$ solutions---that is solutions with scalar hairs---for actual astrophysical objects that are currently supposed to be black holes. The goal would be to test the existence of scalar hairs, despite the fact that they are currently not expected at the theoretical level.

%--------------------------------------------------
\subsection{Comment on the vacuum limit in entangled relativity}

Let us note that entangled relativity corresponds to $f(\varphi,\Lm) = f(\varphi)\Lm(e^{\alpha \varphi} g_{\mu \nu})$ like usual dilaton theories, provided that one has $\Lm \neq 0$ and $R \neq 0$  \cite{ludwig:2015pl, minazzoli:2018pr,arruga:2021pr}. This means that while the solutions (\ref{eq:metric}-\ref{eq:beta}) cannot be exact solutions of entangled relativity, they sould be good approximations in the vacuum limit of the theory---that is, when $T_{\mu \nu} \rightarrow 0$ but $T_{\mu \nu} \neq 0$---just as the Schwarzschild metric has been found to be a good approximation of spherical black holes in this limit in \cite{minazzoli:2021bh}.

%--------------------------------------------------
\section{Dilaton action and solutions in the Einstein frame}
\label{sec:EF}

After the conformal transformation  $\tilde g_{\alpha \beta} = e^{-2\varphip} g_{\alpha \beta}$, with $\phi = e^{-2\varphip}$, the action (\ref{eq:sfaction}) reads
\bea \label{eq:actionclER}
%S=\int d^{4} x \sqrt{-g}\left[R-(\nabla \varphi)^{2} + f(\varphi,\Lm) \right],
S=\int d^{4} x \sqrt{-\t g}\left[\t R-2\t g^{\alpha \beta}\partial_\alpha \varphi \partial_\beta \varphi \right . \\
%\left. + \t f(\varphi,\Lm(e^{2 \varphi / \sqrt{3}} \t g_{\mu \nu})) \right], 
\left. +  e^{- \varphi/\sqrt{3}}{\t \L}_m(e^{2 \varphip} \t g_{\mu \nu}) \right],
\eea
where ${\t \L}_m := e^{4 \varphip} \Lm$. It means that it corresponds to
\be \label{eq:fER}
f(\varphi,\t \L_m) = e^{- \varphi/\sqrt{3}}{\t \L}_m(e^{2 \varphip} \t g_{\mu \nu})
\ee
in Eq. (\ref{eq:actioncl}), in which one would have $g_{\alpha \beta} :=\t g_{\alpha \beta}$. One can check that $\delta(\sqrt{-\t g} f)/\delta \varphi=0~ \forall ~\t \L_m=\t T$---or, equivalently, $\Lm = T$ in the field equations in the original frame---such that the scalar-field equation reduces to $\Box \varphi=0~ \forall ~\t \L_m=\t T$ on-shell. This is the property of what has been called \textit{intrinsic decoupling} in \cite{minazzoli:2013pr,minazzoli:2014pr}. Note that one notably has $\Lm = T$ for dust and pure electromagnetic radiation for instance; whereas one has $\Lm \neq T$ for an electric or a magnetic field---notably leading to different charged black-holes with respect to the ones of general relativity \cite{minazzoli:2021bh}.

One can therefore use the external solutions (\ref{eq:metric}-\ref{eq:beta}), that we shall rewrite as follows (for later convenience):
\bea
ds^2 &=& - \left(1-\frac{2m}{\beta r} \right)^{\frac{1-\alpha^2}{1+\alpha^2}} \dd t^2 +\left(1-\frac{2m}{\beta r} \right)^{-\frac{1-\alpha^2}{1+\alpha^2}} \dd r^2 \nonumber \\
&&+ r^2~\left(1-\frac{2m}{\beta r} \right)^{\frac{2\alpha^2}{1+\alpha^2}} \left[\dd \theta^2 + \sin^2 \theta ~\dd \psi^2 \right], \label{eq:metrica}
\eea
and
\be
e^\varphi = \left(1-\frac{2m}{\beta r} \right)^{\frac{\alpha}{1+\alpha^2}} \label{eq:SFma},
\ee
with
\be
\beta = \frac{1-\alpha^2}{1+\alpha^2}.
\ee
There are three possible branches:
\begin{itemize}
\item $\alpha > 0$: which we shall name $\alpha_+$.
\item $\alpha < 0$: which we shall name $\alpha_-$.
\item $\alpha = 0$: which we shall name $\alpha_0$, and which simply is the Schwarzschild solution.
\end{itemize}
In the Einstein frame, the metric solutions for the branches $\alpha_+$ and  $\alpha_-$ are the same, but it is not the case in the original frame, as we shall see in the next section.

As one can see in Eq. (\ref{eq:SFma}), the sign of $\alpha$ gives the direction of the monotonicity of $\varphi$. One therefore deduces that $\alpha_0$ corresponds to sources that are such that $\delta(\sqrt{-\t g} f)/\delta \varphi=0$, $\alpha_+$ corresponds to sources that are such that $\delta(\sqrt{-\t g} f)/\delta \varphi>0$ and $\alpha_-$ corresponds to sources that are such that $\delta(\sqrt{-\t g} f)/\delta \varphi <0$.

%--------------------------------------------------
\section{Solutions in the original frame}
\label{sec:ERlim}

Performing the inverse conformal transformation---$g_{\alpha \beta} = e^{4 \zeta \varphi} \t g_{\alpha \beta}$, with $\phi = e^{-4 \zeta \varphi}$ and $\zeta := 1/(2\sqrt{3})$---in order to get the correponding solutions for the action (\ref{eq:sfaction}), one either gets the usual Schwarzschild solution as limit (when $\alpha = 0$), or the two limits that follow. With the metric given by 
\be
g_{\mu\nu}\dd x^\mu \dd x^\nu=-a\dd t^2+b\dd\rho^2+\rho^2\dd\Omega^2, \label{eq:metricwrho}
\ee
where $\dd\Omega^2 = \dd \theta^2 + \sin^2 \theta ~\dd \psi^2$, one has
\bea
&&a=\left(1-\frac{2m}{\beta r}\right)^{\frac{1-\alpha^{2}+4  \zeta \alpha}{1+\alpha^{2}}},\label{eq:a_oframe}\\
&&b=\left(\frac{\mathrm{d} \rho}{\mathrm{d} r}\right)^{-2}\left(1-\frac{2m}{\beta r}\right)^{\frac{\alpha^{2}-1+4  \zeta \alpha}{1+\alpha^{2}}},\\
&&\rho=r\left(1-\frac{2m}{\beta r}\right)^{\frac{\alpha^{2}+2  \zeta \alpha}{1+\alpha^{2}}},\\
&&\phi=\left(1-\frac{2m}{\beta r}\right)^{-\frac{4 \zeta \alpha}{1+\alpha^{2}}}.\label{eq:phi_oframe}
\eea
One can see that in this frame, the metric are different for the branches $\alpha_+$ and  $\alpha_-$---which correspond to $\alpha > 0$ and $\alpha < 0$ respectively.

%--------------------------------------------------
\section{Comparing to external numerical solutions}

With the metric given by 
\be
g_{\mu\nu}\dd x^\mu \dd x^\nu=-a\dd t^2+b\dd\rho^2+\rho^2\dd\Omega^, 
\ee
the Tolman–Oppenheimer–Volkoff (TOV) equations for the action (\ref{eq:sfaction}) outside an object in the vacuum limit reduce to
\begin{align}
    &\frac{\ddot{\phi}}{\dot{\phi}} = \frac{1}{2}\left[\frac{\dot{b}}{b}-\frac{\dot{a}}{a}-\frac{4}{\rho}\right]\label{A}\\ 
    &\frac{\dot{b}}{b}=\frac{1-b}{\rho}-\frac{\xi}{2}\frac{\dot{a}}{a} \label{B}\\
    &\frac{\dot{a}}{a}=\frac{b-1}{\rho}\left[1+\frac{\xi}{2}\right]^{-1}-2\frac{\xi}{\rho}\left[1+\frac{\xi}{2}\right]^{-1}\label{C}
\end{align}
where $\xi\equiv \rho\dot{\phi}/\phi$. After intrgrating Eq. (\ref{A}), one ends up with
\begin{equation} \label{dotphi}
    \dot{\phi}=\frac{C}{\rho^2}\sqrt{\frac{b}{a}}
\end{equation}
with $C$ a constant of integration. The sign of $C$---which is such that $sign(C)=-sign(\alpha)$ (see Eq. (\ref{eq:phi_oframe}))---completely determines the monotonicity of $\phi$. 

One can check that each of the three vacuum limits discussed in Sec. \ref{sec:ERlim} is indeed solution to the equations (\ref{A}-\ref{C}). It means that the three solution $\alpha_0$, $\alpha_-$ and $\alpha_+$ should match external numerical solution of the TOV equations found in \cite{arruga:2021pr}.

Entangled relativity being parameter free, the three cases can neverthless be shown to correspond to the three different types on-shell matter Lagragangians that one may consider---that is, $\Lm = T$, $\Lm = -\rho$ or $\Lm = P$, given that they correspond to no-source, postive or negative sources\footnote{The source of the scalar-field equation is proportional to $\L_m - T$ in Eq. (\ref{eq:sceq}), such that it is null, positive or negative for $\Lm = T$, $\Lm = -\rho$ or $\Lm = P$ respectively. The corresponding value of $C$ will therefore be null, positive or negative respectively, such that the corresponding value of $\alpha$ will be null, negative or positive respectively.} in the scalar-field equation respectively. [Depending on the type of matter composing the compact object, one may have any of these cases \cite{arruga:2021pr}]. In what follows, we shall show that it is indeed the case, although the three cases actually more generaly depend on whether the source of the scalar field is null, positive or negative respectively---since the various cases are related to the monoticity of $\phi$, as one can see from Eq. (\ref{dotphi}).

The numerical integration is based on our previous work \cite{arruga:2021pr}, in which we assumed a basic polytropic equation of state $P = K \rho^{\gamma}$ for simplicity, with $\gamma=5/3$ and $K=1.475\times 10^{-3} (fm^3/MeV)^{2/3}$.

The code that does the numerical integration and generates all the figure of this manuscript is freely availbale on GitHub \cite{code2}.

%--------------------------------------------------
\subsection{Comparison of $\alpha_0$}

It was found in \cite{arruga:2021pr} that for $\Lm = T$ the TOV solutions in entangled relativity are the same as the ones of general relativity. It means that for $\Lm = T$, the external vacuum limit solution of compact objects is the Schwarzschild metric. Therefore, the case $\alpha_0$ corresponds to compact objects that are made of matter that satisfy $\Lm = T$. Indeed, the extra degree of freedom of entangled relativity with respect to general relativity is not sourced in that case. Whether or not matter can lead to $\Lm = T$ is an ongoing debate \cite{arruga:2021pr,avelino:2018pr,avelino:2018pd}. 

In any case, as already discussed in Sec. \ref{sec:generic}, due the radiation of scalar hair during the collapse into a black hole in scalar-tensor theories \cite{hawking:1972cm,scheel:1995pr}, one expects that the Schwarzschild metric corrspond to black hole solutions, whether or not the Schwarzschild solution also corresponded to the external solution of the initial spherical compact object.

%--------------------------------------------------
\subsection{Comparison of $\alpha_-$}

\begin{figure}
\includegraphics[scale=0.5]{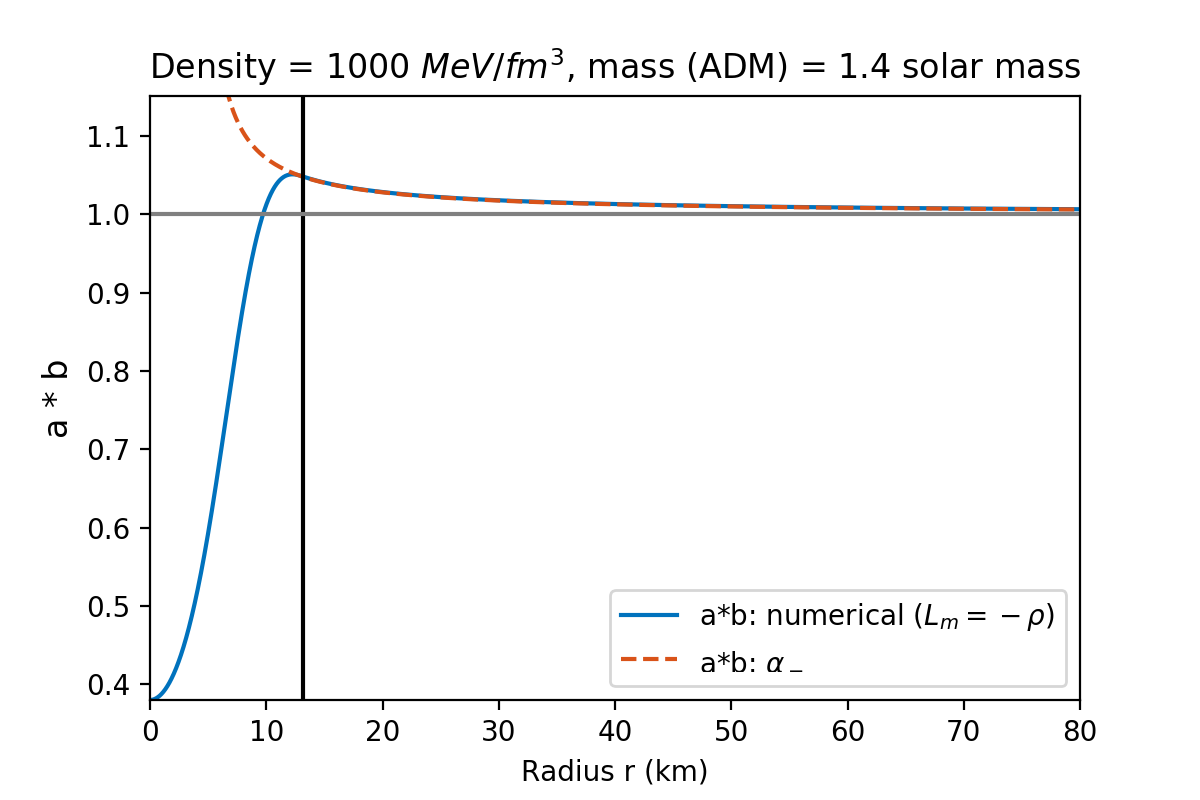}
\includegraphics[scale=0.5]{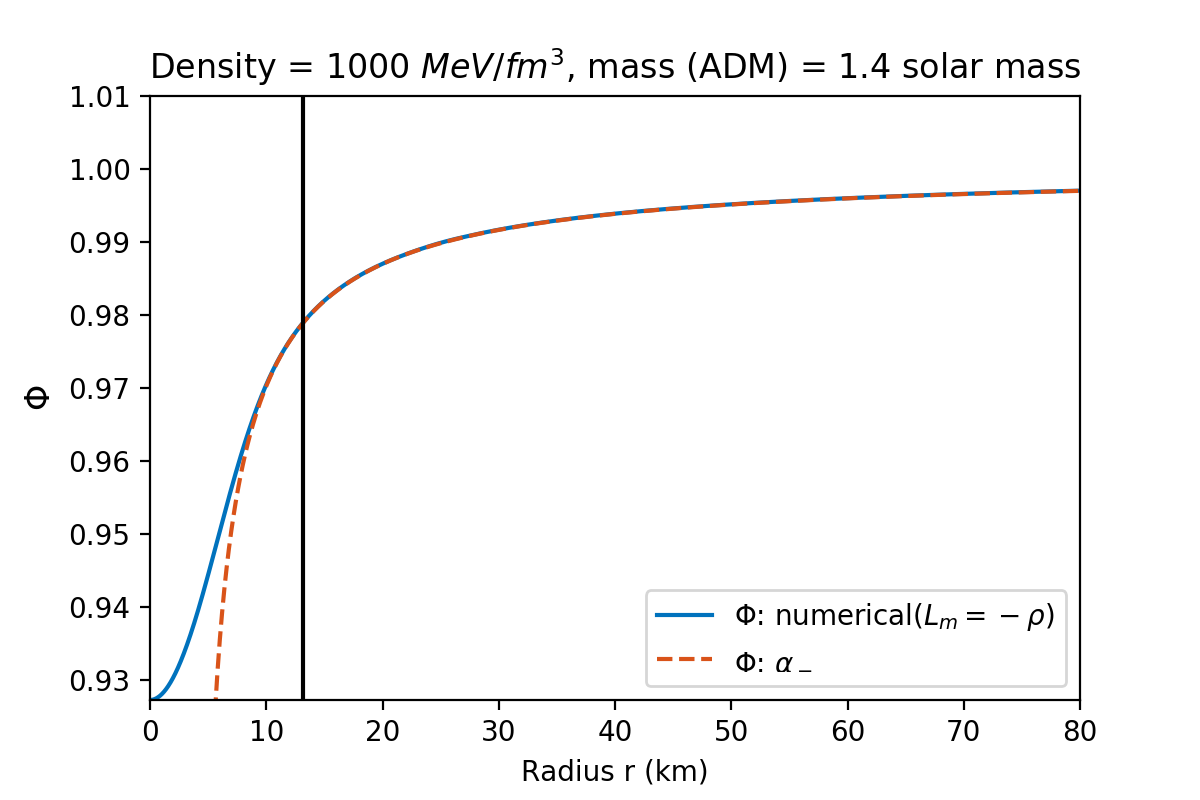}
\caption{Comparison of the product of the metric components and the scalar field between the analytical solution $\alpha_-$ and the numerical solution for $\Lm = - \rho$, and $\rho_0 = 1000~ MeV/fm^3$. The vertical line indicates the radius of the compact object.}
\label{fig:rho}
\end{figure}

\begin{figure}
\includegraphics[scale=0.5]{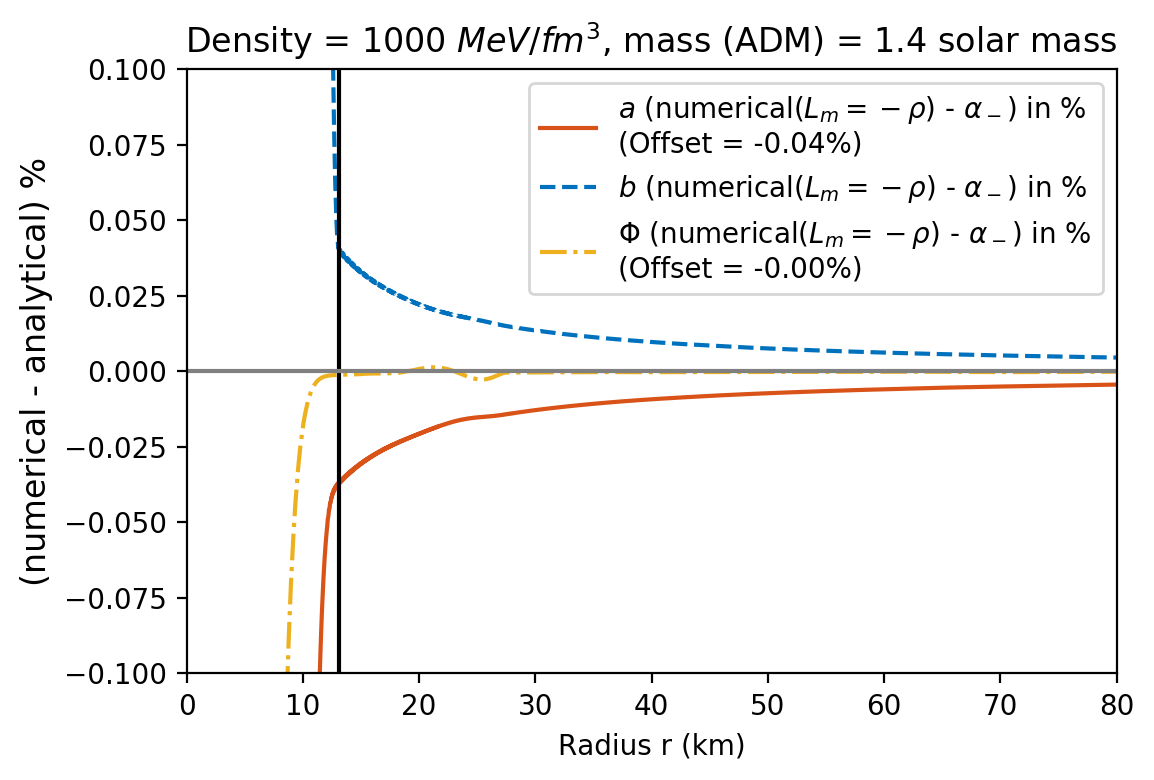}
\caption{\% difference of the metric components and the scalar field between the analytical solution $\alpha_-$ and the numerical solution for $\Lm = - \rho$, for $\rho_0 = 1000~ MeV/fm^3$. The vertical line indicates the radius of the compact object.}
\label{fig:rhodiff}
\end{figure}

The $\alpha_-$ case can be matched to the external part of the numerical TOV solutions for compact objects that are made of matter fields that satisfy $\Lm = -\rho$---as one can see in Figs. \ref{fig:rho} and \ref{fig:rhodiff}. 

In particular, one can see the good agreement between the analytical and external numerical solutions in Fig. \ref{fig:rhodiff}. Only a sub-permil deviation between the two solutions occures at the limit of the compact object, and then decreases with the distance to the object. A sub-permil offset---which has been removed in Fig. \ref{fig:rhodiff}---remains for the time-time component of the metric $a$. It is easily explained by the fact that at the numerical level, one cannot use a normalization at infinity, but only at the limit of the simulation---that is, in our case, at $r=r_\infty:=10.000$ km.

\begin{figure}
\includegraphics[scale=0.5]{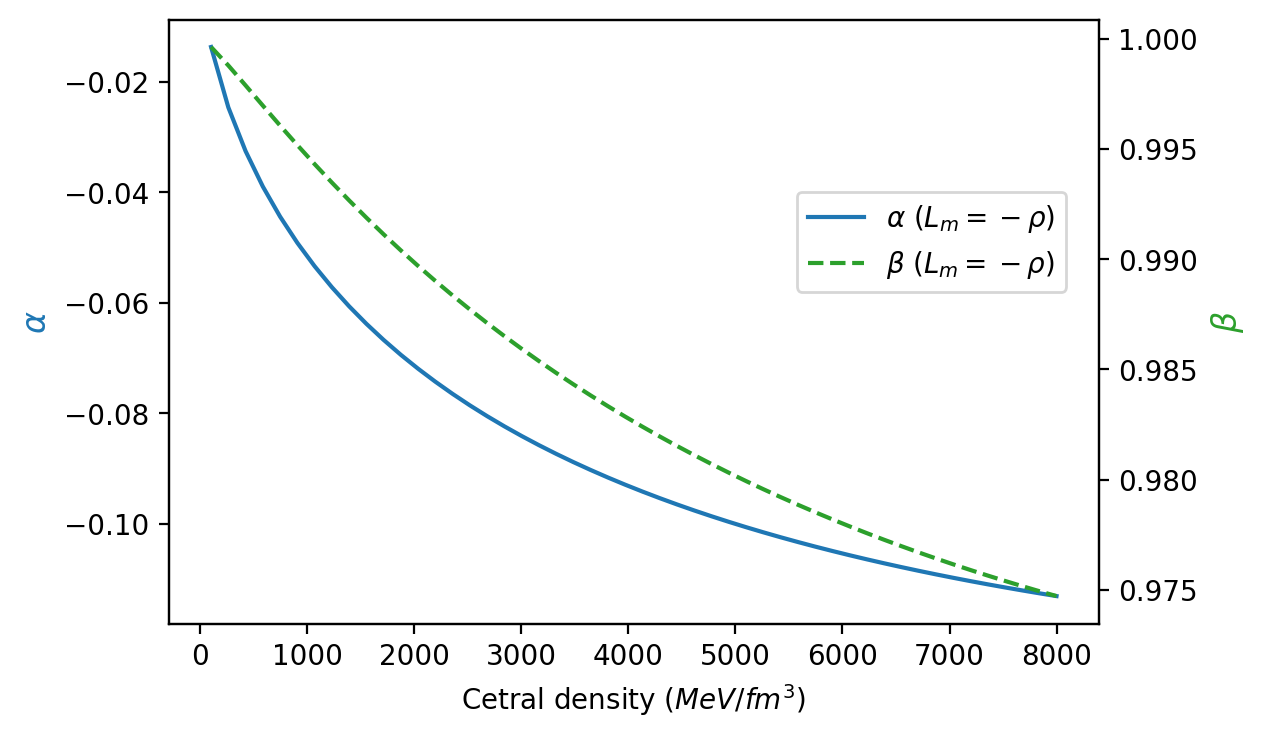}
\caption{Values of the parameters $\alpha$ and $\beta$ in Eq. (\ref{eq:a_oframe}-\ref{eq:phi_oframe}) with respect to the density of the compact object for $\Lm = - \rho$.}
\label{fig:alpha_beta_rho}
\end{figure}

One can see in Fig. \ref{fig:alpha_beta_rho}
% the Appendix \ref{app:gamma_beta}, 
there is a monotonic behavior of the parameters $\alpha$ and $\beta$ in (\ref{eq:a_oframe}-\ref{eq:phi_oframe}) with respect to the central density. For objects with low central densities, the solution is closer to the Scharzschild solution---that is $\alpha$ is closer to 0, and $\beta$ is closer to 1---whereas the stronger the cental density, the more the solution deviates from the Scharzschild solution of general relativity. In other words, the more relativistic the object, the more deviation there is from general relativity.

 More generally, the $\alpha_-$ case corresponds to spherical solutions for which the source of the scalar-field (i.e. r.h.s. of Eq.(\ref{eq:sceq})) is positive.

%--------------------------------------------------
\subsection{Comparison of $\alpha_+$}

\begin{figure}
\includegraphics[scale=0.5]{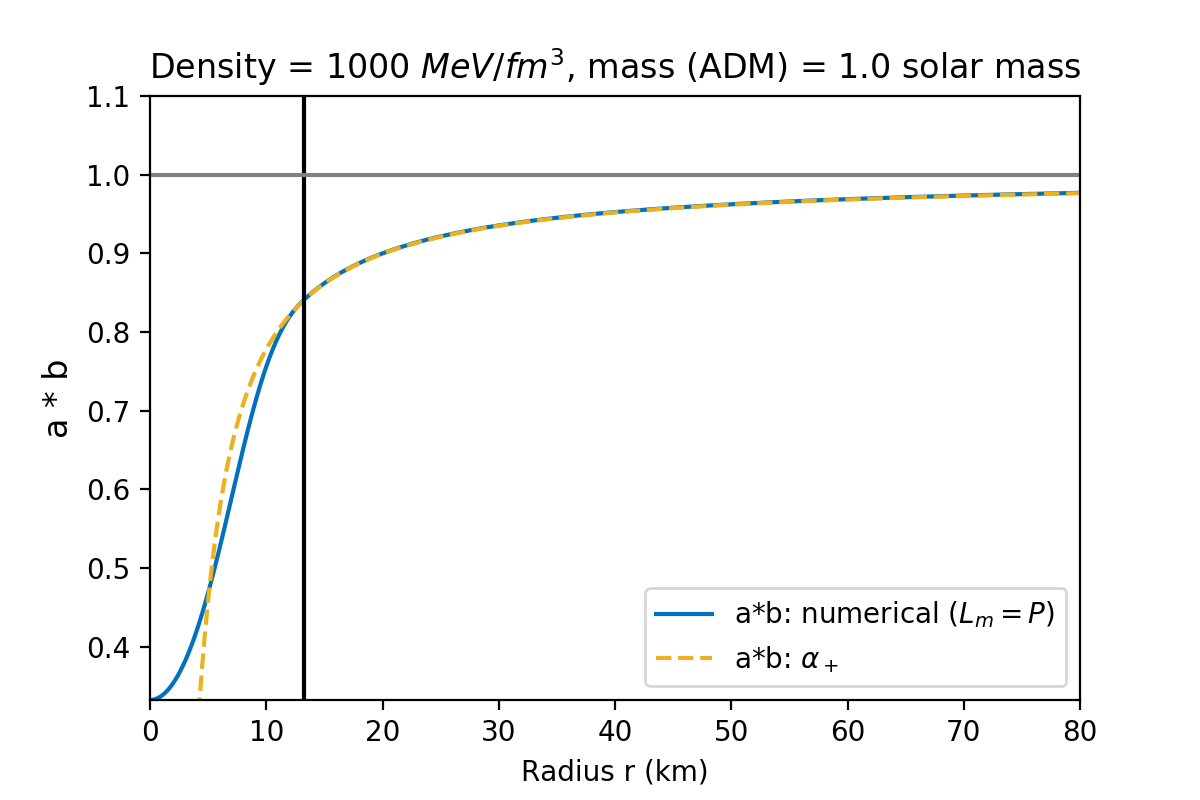}
\includegraphics[scale=0.5]{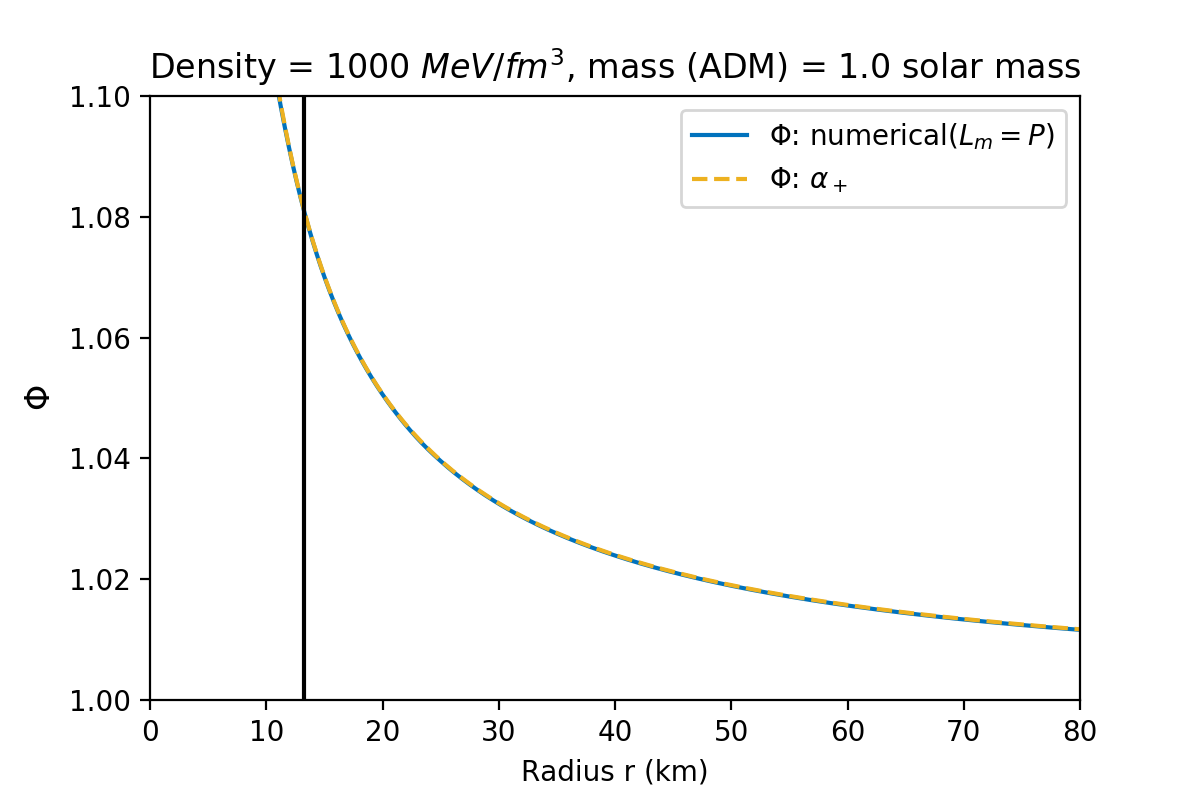}
\caption{Comparison of the product of the metric components and the scalar field between the analytical solution $\alpha_+$ and the numerical solution for $\Lm =P$, and $\rho_0 = 1000~ MeV/fm^3$. The vertical line indicates the radius of the compact object.}
\label{fig:P}
\end{figure}

\begin{figure}
\includegraphics[scale=0.5]{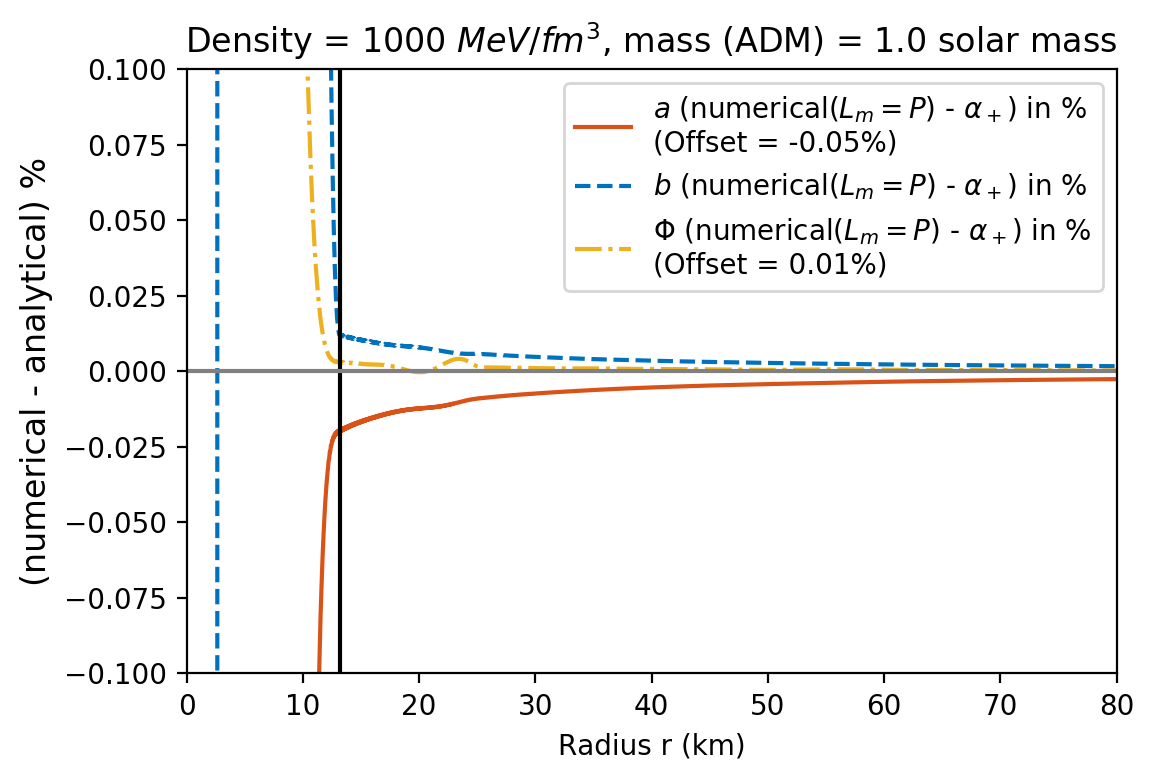}
\caption{\% difference of the metric components and the scalar field between the analytical solution $\alpha_+$ and the numerical solution for $\Lm = P$, for $\rho_0 = 1000~ MeV/fm^3$. The vertical line indicates the radius of the compact object.}
\label{fig:Pdiff}
\end{figure}

The $\alpha_+$ case can be matched to numerical TOV solutions for compact objects that are made of matter that satisfy $\Lm = P$---as one can see in Figs. \ref{fig:P} and \ref{fig:Pdiff}. Although note that $\Lm=P$ is not consistent with a collection of baryonic particles, which Lagrangian must tend to $\Lm = - \rho_0$ for $P=0$, where $\rho_0$ is the inertial energy density of the collection of particles. $\Lm = P$ might be used, however, in order to model exotic objects that would, for instance, entirely be made of a scalar field---given that $P=K-V=\Lm$ for scalar fields, where $K$ and $V$ are the kinetic and potential energy densities respectively. One may have in mind Higgs monopoles for instance, like in \cite{schlogel:2014pr}.

In particular, one can see the good agreement between the analytical and external numerical solutions in Fig. \ref{fig:Pdiff}. Again, a sub-permil deviation between the two solutions occures at the limit of the compact object, and then decreases with the distance to the object. Again, a sub-permil offset remains for the time-time component of the metric ($a$). 

\begin{figure}
\includegraphics[scale=0.5]{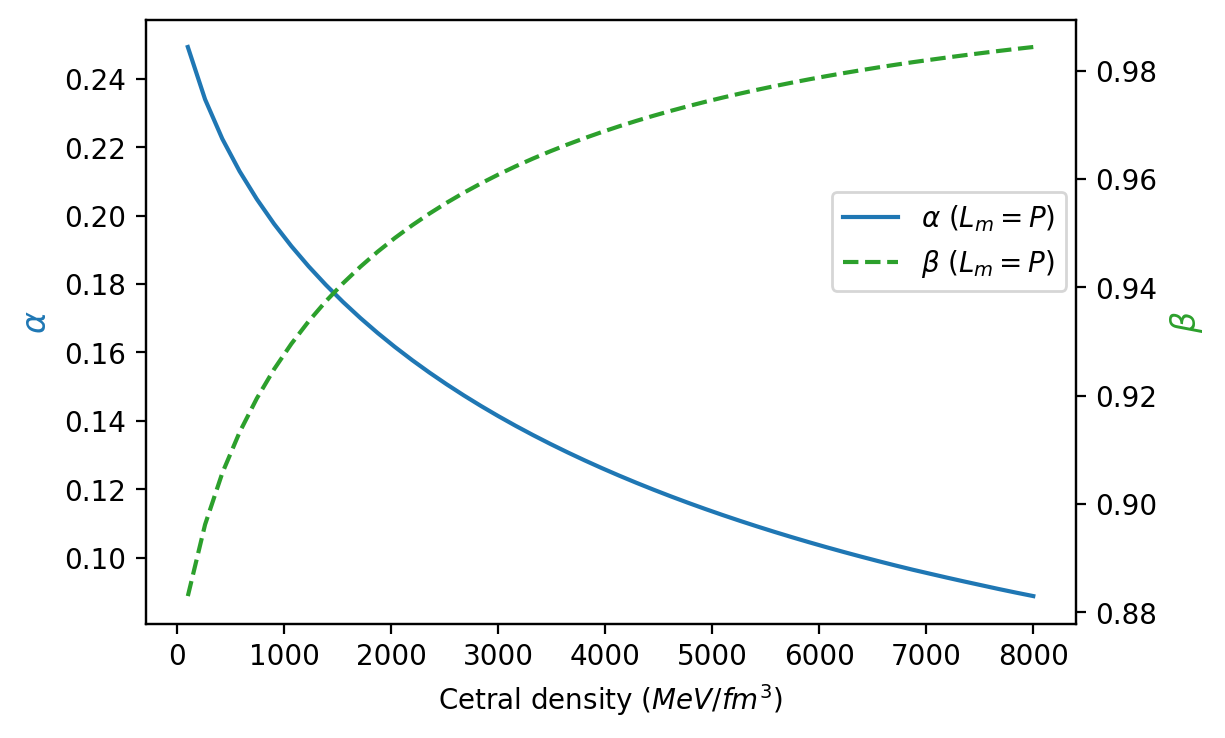}
\caption{Values of the parameters $\alpha$ and $\beta$ in Eq. (\ref{eq:a_oframe}-\ref{eq:phi_oframe}) with respect to the density of the compact object for $\Lm = P$.}
\label{fig:alpha_beta_P}
\end{figure}

As one can see in Fig. (\ref{fig:alpha_beta_P})
%the Appendix \ref{app:gamma_beta}, 
the behavior of the parameters $\alpha$ and $\beta$ in (\ref{eq:a_oframe}-\ref{eq:phi_oframe}) is slightly different for $\alpha_+$. Indeed, the more dense the compact object, the more close the Schwarzschild metric it becomes. [We have checked that this behavior reverses for much lower central densities---that is, that $\beta$ goes back to one for much less dense objects]. %\OMD{However, let us stress again that this solution is not solution of entangled relativity in its original form (\ref{eq:actER})---see Sec. \ref{sec:2sol}.}
Note that, given one has $\Lm = P > 0$, Eq. (\ref{eq:defphi}) implies that $R<0$ for the $\alpha_+$ case. We have checked that it is indeed the case in our TOV simulation.

More generally, the $\alpha_+$ case corresponds to pherical solutions for which the source of the scalar-field (i.e. r.h.s. of Eq.(\ref{eq:sceq})) is negative.

%--------------------------------------------------
%\subsection{Test particle geodesics}
\section{Test particle geodesics}

We assume that the action (\ref{eq:sfaction}) induces for test particles that their action reads $S_\textrm{tp} = - mc^2~\int \sqrt{\phi} d\tau$, where $\tau$ is an affine parameter of the test particle's trajectory defined such that $d \tau = \sqrt{-g_{\alpha \beta} dx^\alpha dx^\beta}$. This assumption follows the assumption of the stability of the universal coupling at the quantum field theory level, such that all the contributions to the mass of a particle are proportional to the same function of the scalar field $\phi$, which can therefore be factorized out as $m(\phi) \rightarrow \sqrt{\phi}~m$. In particular, it means that we assume that 
\be
\sqrt{\phi} \Lm^{SM} \rightarrow \sqrt{\phi} T^\textrm{SM}_\textrm{anomaly}, \label{eq:anomaly}
\ee 
where $\Lm^{SM}$ is the effective low energy limit of the standard model of particles and $T^\textrm{SM}_\textrm{anomaly}$ its corresponding quantum trace anomaly \cite{nitti:2012pr,minazzoli:2016pr}---which gives their mass to composite particles in the standard model of particles \cite{donoghue:2014bk}.

After the conformal transformation $\t g_{\alpha \beta} = e^{-2\varphip} g_{\alpha \beta}$, with $\phi = e^{-2\varphip}$, defined in Sec. \ref{sec:EF}, the test particle part of the action reads $S_\textrm{tp} = - mc^2~\int  d \t \tau$, where $d \t \tau = \sqrt{-\t g_{\alpha \beta} dx^\alpha dx^\beta}$. The whole action with a neutral massive test particle in the conformal frame therefore reads
\bea
\label{eq:dilaton3}
%S=&\int d^{4} x \sqrt{-g}\left[R-2(\nabla \varphi)^{2}+e^{-2 \alpha \varphi} F^{2}\right] \nonumber\\
%&- mc^2 \int d\tau.
S=&&\int d^{4} x \sqrt{-\t g}\left[\t R-2(\t g^{\alpha \beta} \partial_\alpha  \varphi \partial_\beta \varphi)-e^{- \varphip} \frac{\t  F^{2}}{4}\right] \nonumber\\
&&- mc^2 \int d \t \tau,
\eea
where $\tilde F^2 = F^2 := F_{\alpha \beta} F^{\alpha \beta}$ due to the conformal invariance of the electromagnetic part of the action at the classical level, with $F_{\alpha \beta} = \nabla_\alpha A_\beta - \nabla_\beta A_\alpha$, and $A^\alpha$ the electromagnetic four-vector. 

The electromagnetic contribution to the mass $m$ (through the trace anomaly in Eq. (\ref{eq:anomaly})), by definition, breaks the conformal invariance of the electromagnetic part of the action (\ref{eq:sfaction}), leading to the universal effective coupling $S_\textrm{tp} = - mc^2~\int \sqrt{\phi} d\tau$ in the original frame (\ref{eq:sfaction})---which effectively corresponds to having a mass that is proportional to $\sqrt{\phi}$. This is why $m$ can become independent of the scalar-field in the Einstein frame. It is important to stress that---except for the special case of general relativity---it would not be the case (at least, in general) for other functions $f(\varphi,\t \L_m)$ than the one of entangled relativity, given in Eq. (\ref{eq:fER}). This shows how, despite its very unusual form in Eq. (\ref{eq:actER}), entangled relativity actually is pretty close to general relativity, as one can see in Eq. (\ref{eq:dilaton3}). Only a coupling of an effective scalar degree of freedom to the electromagnetic part of the action remains in the Einstein frame for neutral point particles.

Therefore, neutral massive test particles follow geodesics of the conformal metric $\t g_{\alpha \beta}$. However, it is important to stress that the affine parameter on the geodesic $\t \tau$ is not the time given by, say, an atomic clock along the geodesics, since the latter depends on the variation of the fine structure constant that is proportional to $e^{\varphip}$ in this model \cite{minazzoli:2014pr,hees:2014pr}. Entangled relativity therefore breaks the \textit{local position invariance} in general---that is, as long as $\varphi$ is not constant. Fortunately enough, the embedded \textit{intrinsic decoupling} in entangled relativity implies a near constant scalar field in most situations \cite{minazzoli:2013pr,minazzoli:2014pr,minazzoli:2014pl,minazzoli:2020ds,minazzoli:2020bh,arruga:2021pr}.

One can show that electromagnetic plane-waves in the geometric optic approximation also follow null-geodesics of either the metric $g_{\alpha \beta}$ or its conformally transformed version $\t g_{\alpha \beta}$ \cite{minazzoli:2014pr}. This is due to the conformal invariance of the electromagnetic part of the action at the classical level.

Working in the plane $\theta = \pi/2$ without loss of generality, one deduces the following equations of motion from Eq. (\ref{eq:metric})
\bea
%&&\left(\frac{\dd r}{\dd \tau} \right)^2+\left(1-\frac{2m}{\beta r}\right)^\frac{1-\alpha^2}{1+\alpha^2} \left[\epsilon+ \frac{L^2}{r^2}\left(1-\frac{2m}{\beta r}\right)^{\frac{-2\alpha^2}{1+\alpha^2}} \right] = E^2,  \\
&&\left(\frac{\dd r}{\dd \t \tau} \right)^2+\t \lambda^2(r) \left[\epsilon+ \frac{L^2}{\tilde{\rho}^2(r)} \right] = E^2, \\
%&&r^2\left(1-\frac{2m}{\beta r}\right)^{\frac{2\alpha^2}{1+\alpha^2}}\frac{\dd \psi}{\dd \tau} = L,\\
&&\tilde{\rho}^2(r) \frac{\dd \psi}{\dd \t \tau} = L,\\
%&&\left(1-\frac{2m}{\beta r}\right)^\frac{1-\alpha^2}{1+\alpha^2} \frac{\dd t}{\dd \tau} = E,
&&\t \lambda^2(r) \frac{\dd t}{\dd \t \tau} = E,
\eea
where
\bea
&\tilde{\rho}^2(r)=r^2\left(1-\frac{2m}{\beta r}\right)^{1-\beta}, \\
%%&\tilde{\rho}^2(r)=r^2\left(1-\frac{r_-}{r}\right)^{\frac{2\alpha^2}{1+\alpha^2}}, \\
%&\tilde{\rho}^2(r)=r^2\left(1-\frac{2m}{\beta r}\right)^{\frac{2\alpha^2}{1+\alpha^2}}, \\
% & \lambda^2(r) =\left(1-\frac{r_-}{r}\right)^\frac{1-\alpha^2}{1+\alpha^2}.\\
% & \lambda^2(r) =\left(1-\frac{2m}{\beta r}\right)^\frac{1-\alpha^2}{1+\alpha^2}.\\
 &\t \lambda^2(r) =\left(1-\frac{2m}{\beta r}\right)^\beta. \label{eq:tlam}
\eea
and where $E$ and $L$ are the conserved energy and momemtum along the trajectories respectively, and where $\epsilon$ is either equal to $0$ or $1$ for null and time-like geodesics respectively.

\begin{figure}
\includegraphics[scale=0.6]{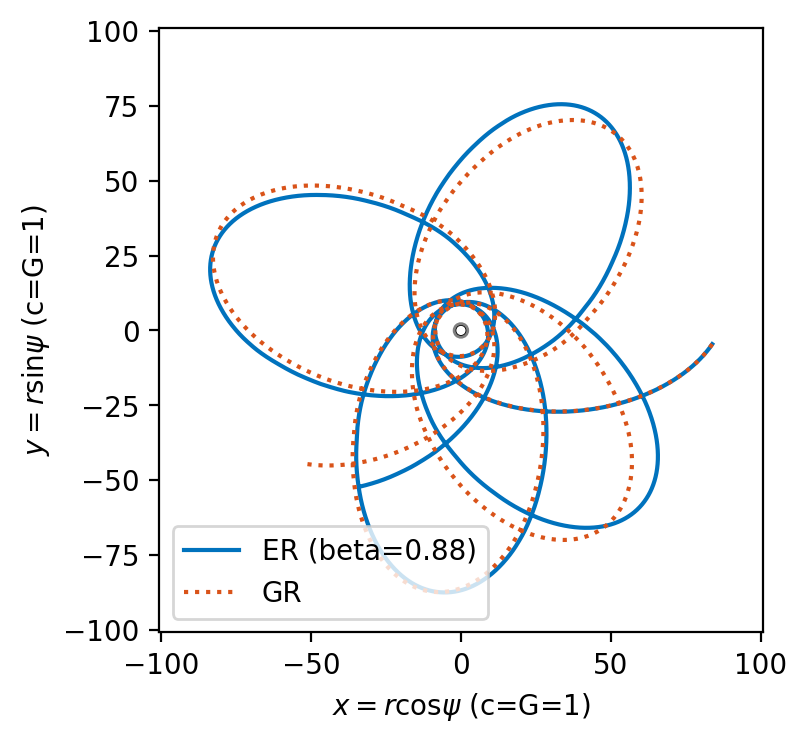}
\includegraphics[scale=0.6]{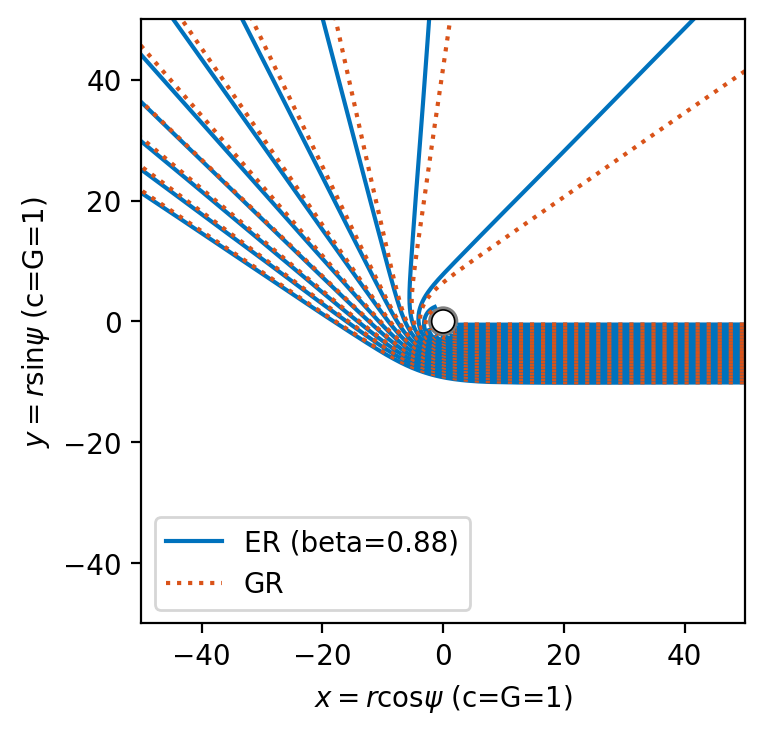}
\caption{Upper panel: timelike geodesics parametrized by the coordinate time $t$, with $m=1$, $E=0.98$, $L = 4.5$, for $\beta =1$ (general relativity) versus $\beta = 0.88$, which corresponds to the lowest value found in our TOV simulations in entangled relativity, for an exotic compact object (i.e. $\Lm = P$)---see Fig. \ref{fig:alpha_beta_P}. Note that for a more realistic object (i.e. $\Lm = - \rho$ or $T$), the lowest value has been found to be $\beta = 0.975$ instead---see Fig \ref{fig:alpha_beta_rho}. Lower panel: null geodesics parametrized by the coordinate time $t$ with various impact parameters,  with $m=1$, for $\beta =1$ (general relativity) versus $\beta = 0.88$.}
\label{fig:time_geod}
\end{figure}

One can define an effective radial potential that reads
\be
V_\textrm{eff}(r) = \frac{1}{2} \left(-E^2 +\t  l^2(r) \frac{L^2}{r^2} +\t \lambda^2(r) \epsilon \right),
\ee
where 
\be
\t l^2(r) := \left(1- \frac{2m}{\beta r} \right)^{2\beta-1}.\label{eq:tl}
%\t l^2(r) := \left(1- \frac{2m}{\beta r} \right)^{1-2\beta}.\label{eq:tl}
\ee
We have plotted some examples of null and timelike geodesics in Fig. \ref{fig:time_geod}.

%--------------------------------------------------
\subsection{Shapiro delay}

The electromagnetic Lagrangian density being conformally invariant at the classical level, one can compute the Shapiro delay from any metric related by a conformal transformation and get the same result. From Eq. (\ref{eq:metric}), the relative time delay \cite{xu:2020pr} from the surface of the spherical object $r=R>2m/\beta$ therefore reads
\be
%\delta t(\sigma) = \int^\infty_{R} \sqrt{\frac{b}{a}}\left(\frac{1}{\sqrt{1-\frac{\sigma^{2} a}{\rho^{2}(r)}}}-1\right) \frac{\dd \rho}{\dd r} ~\dd r%\left(\frac{1}{\sqrt{1-\frac{\sigma^{2} \lambda}{r^{2}}}}-1\right) d r,
\delta t(\sigma) = \int^\infty_{R} \frac{1}{\t \lambda^2(r)}\left(\frac{1}{\sqrt{1-\frac{\sigma^{2} \t l^2(r)}{r^2}}}-1\right) ~\dd r%\left(\frac{1}{\sqrt{1-\frac{\sigma^{2} \lambda}{r^{2}}}}-1\right) d r,
\ee
where the impact parameter $\sigma$ satisfies $\sigma = L/E$, while $\t \lambda(r)$ and $\t l(r)$ are given in Eqs. (\ref{eq:tlam}) and (\ref{eq:tl}) respectively.

$\delta t(\sigma)$ is the traveling time $t(\sigma)-t$, minus the reception time of an hypothetical radial ray $t(\sigma =0)$. It is simply meant to define a usefull finite quantity after an integration over infinity, which can also be used to relate the emission and reception times in terms of the periodic change of the impact parameter of the hot spots at the surface of the neutron star \cite{xu:2020pr}.

%----------------------------------
\section{Conclusion}

In this manuscript, we provided external analytical solutions for spherical objects valid for a very general class of scalar-tensor theories. From them, we derived the specific external analytical solutions for spherical objects in entangled relativity, which we compared to numerical solutions. We found that analytical and numerical solutions match very well outside the compact objects---therefore providing evidence that both types of solutions are valid. 

A direct use of these simple analytical solutions would be to use them instead of the more complex numerical solutions in order to infer, say, a neutron star's mass, radius and scalar hair. Indeed, whereas numerical solutions rely on the unknown equation of state of neutron stars, analytical solutions are parametrized by only two parameters---that is, the mass and a parameter related to the amplitude of the scalar hair. It therefore greatly simplifies the model to be adjusted to observations. 

We also provided the relevant equations in order to compute such observables.

\begin{acknowledgements}
O.M. acknowledges support from the \textit{Fondation des fr\`eres Louis et Max Principale}. The authors thank Alessandro Rousel---owner of the YouTube channel ScienceClic \cite{scienceclic}---for simulating the gravitational lens from the metric in Eq. (\ref{eq:metric}), and for understanding that two metrics with a same values for $m$ and $|\beta|$ were actually the same metric---see Sec. \ref{sec:beta}.
\end{acknowledgements}

\bibliography{ext_star}

\appendix

%------------------------------------
\section{General metric field equation of entangled relativity}
\label{app:fe}

The metric field equation that derives from the action (\ref{eq:actER}), reads ($\forall \Lm$)
\bea
\frac{\Lm^2}{R^2}\left(R_{\mu \nu}-\frac{1}{2} g_{\mu \nu} R\right)&&= - \frac{\Lm}{R} T_{\mu \nu} \nonumber \\
&&+ \left(\nabla_\mu \nabla_\nu - g_{\mu \nu} \Box \right)\frac{\L_m^2}{R^2}. \label{eq:fRmetricfield_o}
\eea
Therefore, for $\Lm \neq 0$, Eq. (\ref{eq:fRmetricfield_o}) can indeed be rewritten as Eq. (\ref{eq:fRmetricfield}).

If the on-shell matter Lagrangian goes to zero during a transition between $\Lm > 0$ and $\Lm <0$ whereas $R\neq 0$, the field equation at the location of the transition reduces to $\left(\nabla_\mu \nabla_\nu - g_{\mu \nu} \Box \right)\L_m^2/R^2 = 0$.

%------------------------------------
\section{Check external analytical solution}
\label{app:equations}
\begin{itemize}
\item From the action in Eq. (\ref{eq:actioncl}), the mertic field equation at the vacuum limit reads
\be
G_{\alpha \beta} = 2 \left[\partial_\alpha \varphi \partial_\beta \varphi - \frac{1}{2} g_{\alpha \beta} (\partial \varphi)^2  \right] =: S_{\alpha \beta},
%&&\Box \varphi = 0,
\ee
where $G_{\alpha \beta}$ is the usual Einstein tensor. From the metric in Eq. (\ref{eq:metric}), one gets the following non-null mixed components of the Einstein tensor
\bea
G^0_0 &=& \frac{\beta^2-1}{\beta^2} \frac{m^2}{r^4} \left(1- \frac{2m}{\beta r} \right)^{\beta-2}\nonumber \\
&=&G^\theta_\theta = G^\psi_\psi = - G^r_r,
\eea
whereas the non-null mixed components of the source reads
\bea
S^0_0 &=& - g^{rr} \left(\frac{\partial \varphi}{\partial r} \right)^2 \nonumber\\
&=& S^\theta_\theta = S^\psi_\psi = - S^r_r.
\eea
One finds that $G^0_0 = S^0_0$, such that one has checked that $G_{\alpha \beta}=S_{\alpha \beta}$.

\item From the action in Eq. (\ref{eq:actioncl}), the scalar field equation at the vacuum limit reads
\be
\Box \varphi = g^{rr} \varphi'' - g^{\alpha \beta} \Gamma^r_{\alpha \beta} \varphi'= 0.
\ee
From the mertic (\ref{eq:metric}) and the scalar field (\ref{eq:SFm}) equations, one gets
\be
g^{\alpha \beta} \Gamma^r_{\alpha \beta} \varphi' = \frac{\beta-1}{\alpha} \frac{2m^2}{r^4} \left(1-\frac{2m}{\beta r} \right)^{\beta-2} \left(\frac{m\beta-r}{m\beta} \right),
\ee
such that one can check that
\be
g^{rr} \varphi'' = g^{\alpha \beta} \Gamma^r_{\alpha \beta} \varphi'.
\ee
Therefore, one indeed has verified that $\Box \varphi = 0$.
\end{itemize}

%------------------------------------
\section{Mapping between the $\beta$ and $-\beta$ solutions}
\label{sec:beta}

Defining $\rho = r - 2m/\beta$, one gets
\be
1-\frac{2m}{\beta r} = \left(1+\frac{2m}{\beta \rho} \right)^{-1},
\ee
such that the equations (\ref{eq:metric}-\ref{eq:SFm}) read
\bea
ds^2 &=& - \left(1+\frac{2m}{\beta \rho} \right)^{-\beta} \dd t^2 +\left(1+\frac{2m}{\beta \rho} \right)^{\beta} \dd \rho^2 \nonumber \\
&&+ \rho^2~\left(1+\frac{2m}{\beta \rho} \right)^{1+\beta} \left[\dd \theta^2 + \sin^2 \theta ~\dd \psi^2 \right], \label{eq:metricp}
\eea
that is, it is Eq. (\ref{eq:metric}) with $\beta \rightarrow - \beta$ when $r \rightarrow \rho$. In particular, one can see that the metric in Eq. (\ref{eq:metric}) with $\beta = -1$ corresponds to the exterior of the Schwarzschild solution---that is, $\beta =1$ in Eq. (\ref{eq:metric})---because $\rho$ goes from the event horizon $r=2m$ to infinity.
Otherwise, the scalar field transforms as
\be
%e^\varphi = \left(1-\frac{2m}{\beta r} \right)^{\frac{1-\beta}{2\alpha}} \label{eq:SFm}
\varphi = \frac{\beta-1}{2\alpha}~\ln \left(1+\frac{2m}{\beta \rho} \right) \label{eq:SFp}.
\ee
%------------------------------------
\section{Curvature singularity}
\label{app:sing}

The Ricci scalar of the metric in Eq. (\ref{eq:metric}) reads
\be
R = -2 \frac{\beta^2-1}{\beta^2} \frac{m^2}{r^{4}} \frac{1}{\left(1 - \frac{2 m}{\beta r} \right)^{2-\beta}},\label{eq:ricci}
\ee
such that one can check that it diverges at $r = 2m/\beta~ \forall~ \beta \in ~]0; 1[$. $r = 2m/\beta$ therefore is a curvature singularity for all $\beta \in~]0; 1[$. For $\beta \in ~]-1;0[$, the Ricci scalar only diverges at $r=0$. The reason simply being that the metric with $\beta < 0$ correspond to the metric with $\beta > 0$ with a shifted radial coordinate $r \rightarrow r + 2m/\beta$. See Sec. \ref{sec:beta}.
%, as in the Schwarzschild case. 
For $|\beta| < 1$, the Ricci scalar is positive, whereas one recovers the flatness of the Schwarzschild metric---that is $R=0$---for $|\beta| = 1$

\end{document}